\documentclass[a4paper,11pt]{article}

\usepackage[utf8x]{inputenc}
\usepackage[top=3.3cm, left=2.5cm, right=2.5cm, bottom=3cm]{geometry}
\usepackage{graphicx}
\usepackage{amsmath}
\usepackage{amsfonts}
\RequirePackage{amsthm,amsmath,amsfonts,amssymb,bbm,mathtools}
\RequirePackage[round,authoryear]{natbib}
\usepackage{authblk}
\usepackage{comment}
\usepackage{url}

\newcommand{\E}{\mathbb{E}}

\begin{document}

\title{What you see is not what is there: Mechanisms, models, and methods for point pattern deviations}

\author[1]{Peter Guttorp}
\author[2]{Janine Illian}
\author[3]{Joel Kostensalo}
\author[4]{Mikko Kuronen}
\author[4]{Mari Myllym\"aki}
\author[5]{Aila S\"arkk\"a}
\author[1,6]{Thordis L. Thorarinsdottir}
\affil[1]{Norwegian Computing Center, Oslo, Norway}
\affil[2]{School of Mathematics and Statistics, University of Glasgow, Glasgow, Scotland, UK}
\affil[3]{Natural Research Institute Finland (Luke), Joensuu, Finland}
\affil[4]{Natural Research Institute Finland (Luke), Helsinki, Finland}
\affil[5]{Department of Mathematical Sciences, Chalmers University of Technology and University of Gothenburg, Gothenburg, Sweden}
\affil[6]{Department of Mathematics, University of Oslo, Oslo, Norway}

\date{}

\maketitle

\begin{abstract}
\noindent
Many natural systems are observed as point patterns in time, space, or space and time. Examples include plant and cellular systems, animal colonies, earthquakes, and wildfires. In practice the locations of the points are not always observed correctly. However, in the point process literature, there has been relatively scant attention paid to the issue of errors in the location of points. In this paper, we discuss how the observed point pattern may deviate from the actual point pattern and review methods and models that exist to handle such deviations. The discussion is supplemented with several scientific illustrations. \\

\end{abstract}



\section{Introduction}
Point process models are a modeling framework for stochastic processes that consist of point occurrences in some space, typically a Euclidean space. Historically, point processes emerged from scientific problems related to, for example, astronomy, telecommunication, or life tables \citep{GuttorpThorarinsdottir2012,DaleyVere-Jones2003b}. More recently, point process models have been commonly used for spatial data analysis \citep{MoellerWaagepetersen2007, Illian&2008, Baddeley&2015} with applications in such diverse fields as forestry \citep{StoyanPenttinen2000}, hydrology \citep{Rodriguez-Iturbe}, ecology \citep{Law&2009, zhang2023}, oncology \citep{Jones-Todd2019} or seismology \citep{KaganKnopoff1980}.

At a basic level, the data are given as point patterns---a set of point coordinates describing the locations of objects or events. The data analysis generally aims to infer aspects of the joint distribution of the point locations; this paper focuses on the statistical analysis of such point patterns. Additionally, each point may be associated with a mark describing further characteristics such as size or type of object. While point pattern analysis has some commonalities with more classical statistical analysis, it is in many ways distinct. 

As for most other data, observed point patterns are subject to both systematic and random observational error. The observed point pattern may thus deviate from the true, underlying point pattern of interest. However, modeling frameworks discussed in  textbooks on the statistical analysis of point patterns lack explicit treatment of observational error. Rather, we have found that such discussion is sparsely scattered across diverse strands of literature, and often aimed at a specific application. The goal of this review paper is to provide a unified overview of the literature on the topic of modeling discrepant point patterns with a specific focus on identifying connections between methodological developments aimed at diverse applications. 

There are two main approaches to dealing with discrepant observations of a point process. As one approach, one can model the full process, i.e., the process generating both the original points and the difference between the original points and the observed ones. However, for many point process models it is difficult to write down the likelihood of the point process of interest without taking discrepancies into account. Adding a modeling component for the discrepancies will be even more challenging, although it is sometimes feasible. As a simple example, adding i.i.d.\ measurement errors to the locations of a homogeneous Poisson process yields another homogeneous Poisson process of the same intensity. Another possibility is to study, analytically or by simulation, the effect of discrepancies on parameters and/or parameter estimates in a point process model. This paper provides examples of both approaches.

Since data discrepancies are closely linked to observational mechanisms, substantial space is given here to a review of relevant observational mechanisms for point processes and potential causes of discrepancies related to these. Similarly, we aim to bring together diverse applications in which discrepant observations are common. 

This review is certainly not all-encompassing and the included material inadvertently somewhat reflects the authors' interests and expertise. In our literature review, we initiated searches in Google Scholar and Web of Science using the terms "point process error", "location error", "measurement error in point processes" and several variants of these. Furthermore, we searched forward through the citations of the papers we found as well as in particular scientific applications.

The remainder of this paper is organized as follows. Section 2 provides a brief introduction to point processes. Modeling frameworks that adjust for observation errors are reviewed in Section 3. The subsequent Sections 4--8 provide a review of several applications where the observation mechanisms for the point patterns cause discrepancies between actual and observed patterns. Some of these applications include unpublished material for which we show illustrative plots. We end the paper with a discussion and suggestions for future work on this topic in Section 9. 

\section{Point process preliminaries}

A spatial point process $X$ is defined as a random set of locations in some spatial domain of interest, here assumed to be $\mathbb{R}^d$ for $d \in \{2,3\}$ or a subset thereof. A point pattern $\mathbf{x} = \{x_1,\ldots,x_n\}$ is a realization of a point process observed in a bounded window $W$. We assume that the number $n(A)$ of points of the process in any bounded set $A$ is finite, and also write $n(\mathbf{x})$ to denote the number of points in the observed point pattern $\mathbf{x}$. A spatial point process is called stationary if the distribution of the process is invariant under translation, while it is called isotropic if the distribution is invariant under rotation. See, e.g., \cite{Guttorp1995} for a more detailed definition of point a process. 

\subsection{Summary statistics}

In the analysis of point pattern data, summary statistics are widely used for exploratory analysis, for model fitting and to assess the goodness-of-fit of estimated models. 
There are some standard summary statistics of point patterns. The intensity $\lambda (x)$ of a point process is the  expected number of points per unit area (volume). That is, letting $S_{x,\epsilon}$ be a ball of radius $\epsilon$ centered at $x$, 
$$
\lambda (x) = \lim_{\epsilon \rightarrow 0} \frac{\E n(S_{x,\epsilon})}{|S_{x,\epsilon}|}  ,
$$
where $|A|$ denotes the area or volume of the set $A$.
For a stationary process the intensity is constant. A constant intensity can be estimated by 
$$\hat{\lambda} = \frac{ n(W)}{|W|},$$
while for a non-stationary point process, the intensity and other parameters are location dependent. 

The second-order intensity measures the joint intensity at two locations and is given by  
$$
\lambda_2(x,y) = \lim_{\epsilon_1 , \epsilon_2 \rightarrow 0} \frac{ \E \left(n(S_{x,\epsilon_1})n(S_{y,\epsilon_2})\right)}{|S_{x,\epsilon_1}| |S_{y,\epsilon_2}|},
$$
assuming it exists, see, e.g., \cite{Illian&2008} for more details. For a stationary isotropic point process the second-order intensity depends only on the distance $\|x-y\|$ between $x$ and $y$, and we write it as $\lambda_2(\|x-y\|)$. 

For stationary isotropic processes one often considers the $K$-function,
$$K (r) = \frac{2 \pi}{\lambda^{2}} \int_0^r u\lambda_2(u)du.$$
This can be thought of as the expected number of additional points within distance $r$ of an arbitrarily chosen point, normalized by the intensity. In order to estimate $K(r)$ from data, given by occurrences in a window $W$, we count
the number of occurrence pairs of distance at most $r$ apart. The expected value of this
quantity is approximately $\lambda^2 |W| K (r)$, where the approximation arises from the
possibility that an event in $W$ can be within distance $r$ of the boundary of $W$, and
hence have a neighbor falling outside of $W$. While one can correct for this edge effect, one can also simplify matters by ignoring all points that fall within distance $r$ of the boundary. Often  variants of the $K$-function are used, including  the  variance stabilizing  $L$-function $L(r)=\sqrt{K(r)/\pi}$ and the non-cumulative pair-correlation function $g(r)= \frac{K'(r)}{2 \pi r}$, which is essentially the derivative of the $K$-function and may be interpreted as the number of additional points \textit{at} distance $r$.

Another common second order summary statistic (i.e., involving pairs of points) is the nearest neighbor distance function $G$. For each point $x_i$ in the pattern $\mathbf{x}$ we find the distance $r_i$ to the nearest other point in $\mathbf{x}$. The empirical distribution function of the distances is a natural estimate of the theoretical nearest neighbor distribution, which of course depends on the point process model used. If instead, the distance from an arbitrary point in space to the nearest point of the process is of interest, the empirical distribution of empty space distances (or empty space statistic) can be used. 

While the intensity in general does not determine the distribution of the process, the Papangelou conditional intensity provides a full characterization of the process. It is defined as 
$$
\lambda(u|\mathbf{x}) = \left\{\begin{array}{lll}
\frac{f(\mathbf{x}\cup \{u\})}{f(\mathbf{x})},\quad u\notin\mathbf{x}\\

\frac{f(\mathbf{x})}{f(\mathbf{x}\setminus\{u\})},\quad u\in\mathbf{x}
\end{array}\right.
$$
and is essentially the conditional probability of observing a point of the process $\mathbf{X}$ at $u$ given the observed point pattern $\mathbf{x}$ not including $u$.

\subsection{Point process models}

The simplest stochastic model for a point process is a Poisson process, in which the counts in disjoint sets are independent Poisson random variables. If the mean of each Poisson-variable is proportional to the area of the set, the Poisson process is called homogeneous, otherwise non-homogeneous. In the homogeneous case, the conditional distribution of points, given the total number of points in the set, is that of independent uniform random variables. Therefore, a homogeneous Poisson point process is called completely spatially random. Compared to a Poisson process, a point pattern is often considered clustered or regular (although it may be neither, or it may be both at different scales). A common generalization of the non-stationary Poisson process is the doubly stochastic Poisson process (or Cox process), where the intensity function of a non-stationary Poisson process is a realization of a stochastic process. 
 The log Gaussian Cox process, a Cox process where the logarithm of the intensity is a Gaussian process, is popular in applications since it is completely characterized by the intensity and pair correlation function of the Cox process. 

The most common clustered model is the Neyman-Scott process, which consists of a primary point process of cluster centers, and for each cluster center a secondary point process of cluster points, often independently placed around the primary point. The primary process may or may not be observed.

As to regular patterns, perhaps the simplest is a Poisson process from which all pairs of points closer than a distance $d$ are removed. Such a process is called a hard core rejection model. Other common point process models for regular patterns are Gibbs processes given by a density $f$ with respect to a unit rate Poisson process. For a neighborhood relation $\sim$ (describing which points are considered neighbors), the density can be written
$$
f(\mathbf{x}) \propto \prod_{\mathbf{y} \subset \mathbf{x}} \phi(\mathbf{y}),
$$
where the non-negative interaction function $\phi (\mathbf{y}) = 1$ if there is any pair of points $(y_i,y_j)$ in the pattern $\mathbf{y}$ with $y_i \nsim y_j$. A simple nearest neighbor interaction function is of the form $ \gamma ^ {\mathbbm{1} (\|y_i - y_j\| \leq R)} $, where $\mathbbm{1}$ denotes the indicator function and $0 \leq \gamma \leq 1$ measures the strength of inhibition of points within distance $R$. This process is called a Strauss process. In general, it is common to consider only pairwise interactions. 

The historical development of these and related point process models is discussed in \cite{GuttorpThorarinsdottir2012} and technical details can be found in, e.g., \cite{DaleyVere-Jones2003b}, \cite{Lieshout2000}, and \cite{MoellerWaagepetersen2007}. 

\subsection{Statistical inference}

To carry out inference for a point process we need to define a likelihood (technically, the Radon-Nikodym derivative of the point process measure with respect to a unit rate homogeneous Poisson process). The log-likelihood for a non-homogeneous Poisson process with rate $\lambda(x)$ is 
$$
\ln L(x_1,\dots,x_k; \boldsymbol{\theta}) = \sum_{i=1}^k \ln \lambda(x_i; \boldsymbol{\theta}) - \int_W \left(\lambda(x); \boldsymbol{\theta})-1\right)dx.
$$

The joint density or  likelihood function for Gibbs processes includes a normalizing constant that usually cannot be computed analytically, making the likelihood inference computationally expensive. An alternative for the maximum likelihood method is the pseudo-likelihood method, where the likelihood function is approximated by a product of Papangelou conditional intensities which do not include the normalizing constant.

For a more comprehensive review of inference tools for spatial point patterns, see \cite{MoellerWaagepetersen2007},  \cite{Illian&2008} and \cite{Baddeley&2015}.

\section{Models for deviant point patterns}

Observed point patterns may deviate from the actual point pattern in a multitude of ways. There may be \textit{recording errors} in location recordings. For example, the coordinates of points may have been accidentally switched in the field records. A variant is rounding of data due to changes in precision of measurement devices. As an illustration, early Icelandic volcanic eruptions are known only to the nearest year, or locations of trees are recorded only up to a certain accuracy dependent on the measurement device used. More generally, points can be displaced due to \textit{measurement error}. For example, the location of an earthquake is determined using inversion of arrival times of primary waves (P-waves) and secondary waves (S-waves) at several seismometers. The location process itself includes uncertainty, as does each arrival time used in the inversion. 

Points may be lost by \textit{censoring}, for example due to location measurement error pushing the point outside the observation area, but also by points being located too close together to be observed as separate points. Conversely, a \textit{ghost pattern} in the form of points that do not belong to the point pattern of interest may appear. An example might be a pattern formed by landmines, where many mine-like objects are located, and the real land mines must be identified. A second example with similar issues is automatic detection of animals in remote sensing images or through drone surveys, e.g., detection of seal pups on ice floes or orangutan nests in a rainforest \citep{Milne2021}.

\textit{Thinning} occurs when a process separate from that generating the true pattern operates on the pattern by removing points. This can, e.g., occur in transect sampling, where the probability of observing a point can be related to the distance from the transect. Alternatively, a smooth \textit{deformation} of the point pattern may happen. This can, for example, result from optical or instrumental distortion of observations from space.

Below, we review modeling frameworks that have been proposed to account for these various deviations in the observations.

\subsection{A contamination model for point process data}

A fairly general model for \textit{contamination} of point processes is a combination of thinning, ghost pattern, and location error. Following \cite{assuncao1999} we denote the actual point process (or more precisely, its counting measure) by $X$, an independent superposed point process by $Y'$, and a deletion process by $X'$ taking on values 0 or 1, corresponding to deletion or retention, respectively. Writing the contaminated process $Y$ we have
$$
Y(B) = \int_B X'(x) X^{\ast}(dx) + Y'(B).
$$
Here $X^{\ast}$ corresponds to the point locations of $X$ perturbed by measurement error. \cite{LundRudemo2000} assume that $Y'$ is a Poisson process. For a hard-core process $X$, \cite{Redenbach2009} suggest a mechanism for creating outliers violating the hard-core condition by picking a point in the true pattern at random and adding another point (outlier) in the vicinity of this point. 

In presence-only and citizen science data, false absences are accounted for by detection probabilities. To account for false positives, on the other hand, \cite{RoyleLink2006} introduce a mixture model that estimates the probability of detecting an absent species. However, the estimated probability can be inflated by detection heterogeneity, and a reliable estimation of false positives requires a subset of the observations to be unambiguously true detections, see \cite{AltweggNichols2019} and references therein. 

\subsection{Deviant observations conditional on unobserved truth}\label{sec:LundRudemo}
\label{subsec:deviant}

\cite{LundRudemo2000} estimate tree-top locations of Norway spruce ({\em Picea abies}) from an aerial photograph having the candidate tree locations determined either by kernel smoothing \citep{DralleRudemo1996, DralleRudemo1997} or template matching \citep{LarsenRudemo1998} as the starting point.
They consider a pair of point processes, the true process $X=\{X_i: i\in I\}$, $I=\{1,\ldots,n\}$, and the process of imperfect observation $Y=\{Y_i: i\in J\}$, $J=\{1,\dots,m\}$ of $X$. Let $g_{gh}(\cdot)$ be the density for ghost points, $p(\cdot)$ the retention probability, and $g(\cdot|X_i)$ the displacement density. Then, the conditional likelihood of $Y$ given $X$ 
becomes
$$ 
L(Y|X) = \exp\left(|W|-\int_W g_{gh}(y)\,dy\right) \sum_{\mathclap{{\substack{I_1\subseteq I \\ J_1\subseteq J\\|I_1|=|J_1|}}}}\sum\limits_{\pi} L_1L_2L_3,
$$
where the inner sum is over all bijections $\pi:I_1\rightarrow J_1$ for two finite sets $I_1$ and $J_1$ with the same cardinality (denoted by $|\cdot|$). Furthermore,
$$
L_1=\prod\limits_{i\in I_1}p(X_i)g(Y_{\pi(i)}|X_i),
$$
$$
L_2=\prod\limits_{i\in I\setminus I_1}\left(p(X_i)\int_{\mathbb{R}^d\setminus W}g(y|X_i)\, dy+1-p(X_i)\right),
$$
and 
$$
L_3=\prod\limits_{j\in J\setminus J_1}g_{gh}(Y_j).
$$
When both $X$ and $Y$ are observed, the likelihood can be used to obtain information on the disturbance mechanisms that result in the transformation from $X$ to $Y$. \cite{LundRudemo2000} include three disturbance mechanisms in their model: thinning, displacement, and a ghost pattern. The thinning probability is assumed to be constant, i.e., $p(x)=p$. The  original points $X_i$ are displaced, independently of each other, according to a Gaussian probability density, where a non-zero mean corresponds to systematic error. Points displaced outside the observation window are censored. Finally, ghost points are added according to a Poisson process.

\cite{LundEtal1999} suggest a Bayesian analysis in a similar set-up (without censoring). 
Three Gibbs point process models, namely Poisson process, Strauss process, and a process with a logistic pairwise interaction function, are considered as prior distributions for the unobserved true point pattern. 
Given the prior distribution for $X$ and the model parameters, a Metropolis-Hastings sampler is constructed to sample from the posterior distribution of $X$ given $Y$. In a follow-up paper \citep{LundThonnes2004}, the Metropolis-Hastings algorithm is replaced by perfect simulation. 

\subsection{Adjusted intensity estimates}

One common approach to estimating the intensity of a point process model is based on kernel tools. If these are used in the context of a contaminated point process, deconvolution methods need to be applied. \cite{Diggle1985} suggests a kernel  estimator of point process intensity for temporal non-stationary point processes observed with \textit{}t{location errors},  subsequently extended to spatial data by \cite{Cucala2008}.
In the simplest case of no edge correction, the intensity estimate (with $\mathcal{F}$ denoting the Fourier transform and $b$ the bandwidth) is
$$
\lambda_b^\ast (x) = \mathcal{F}^{-1} ( \mathcal{F}(\hat{\lambda}_b )/ ( \mathcal{F}(g) ) (x).
$$
Here, $g$ is the density of the location errors and 
$$
\hat{\lambda}_b (u) = \int k ((u - x)/b) dX (x) /b^2
$$
is a kernel estimate of the intensity of point process $X$ using kernel $k$.

Another scenario concerns point processes where some of the points that are present have not been detected during data collection. This is common in what is referred to as 
\textit{distance sampling} where data are collected along transects used in animal ecology when the area of interest is typically too big to be sampled exhaustively in its entirety. Here, the effect that objects closer to the observer are more likely to be observed than those further away are included in the modeling framework using a detection probability---decreasing in distance from the observer---as a component of the density. That is, the model is given by a thinned point process and hence an adjusted intensity with unknown thinning probability that may evolve in time and space \citep{HedleyBuckland2004, Renner&2015, Yuan&2017}. Similarly, one of the approaches for deriving stand level characteristics from terrestrial laser scanning (TLS) data is based on a distance sampling methodology \citep{DuceyAstrup2013}. 

For a fixed point in time, denote the true spatial process by $X$, and the process of imperfect observation of $X$ by $Y$, as in Section~\ref{sec:LundRudemo} above. Assume the observer is moving at a constant speed $v$ along a line transect and denote the perpendicular distance from the line transect by $z(x)$. Then, 
\begin{align*}
\mathbbm{P}( \text{object at $x$ detected} | x \in X) & = 1 - \exp \Big( - \frac{h (z(x))}{v} \Big) \\
& = g (z(x), v),
\end{align*}
where $h$ is an aggregated detection hazard along a path and $g$ is the aggregated detection function. The intensity of the observed process $Y$ is then given by
$$
\lambda^{\ast} (x) = \mathbbm{P}( \text{object at $x$ detected} | x \in X) \lambda (x),
$$
where $\lambda$ is the intensity of the true process $X$. For computational convenience, the detection probability is commonly parameterized in terms of the aggregated detection function $g$ rather than the aggregated detection hazard $h$, either as a half-normal density or a generalization thereof, see, e.g., \citet{Yuan&2017}. In Section \ref{sec:dist} we discuss distance sampling in some more detail and provide an example of a marked spatial point process observed through distance sampling, where the detection function depends on the mark. 

\cite{ChakratobortyGelfand2010} assume that the true unobserved point pattern is a realization of a non-homoge\-neous Poisson process, where the
intensity is modeled by a scaled Gaussian mixture distribution. The observed point pattern is a result of random perturbation of the true point locations. The perturbation model is $y_i=x_i+\epsilon_i$, where the noise variables $\epsilon_i$ are i.i.d.\ random variables.
Two scenarios are considered: the perturbation may cause a true point moving outside the observation window and being missed or a point outside the window moving inside the window and being falsely counted as a point within the window. Given the observed noisy data, the true intensity function is estimated in a hierarchical Bayesian framework, where the unknown point pattern is modeled by a Cox process. The true number of points is one of the model parameters. They apply their methodology to estimate the intensity of ecological field data in the Cape Floristic region in South Africa.

Capture-recapture settings in ecology commonly require adjusted intensity estimates. In this context, \cite{Chandler&2011} suggest a hierarchical model for inferring the density of unmarked populations--that is, data where the identity of an individual is not known--subject to temporary emigration and imperfect detection. 

In an application to geocoded spatial data, \cite{FanshaweDiggle2011} propose a general model-based solution to correct for positional errors in various types of spatial data. Specifically, they derive the likelihood function for a linear Gaussian geostatistical model incorporating positional errors. However, the inference method is not computationally feasible for even moderately large data sets. In a follow-up work, \cite{Fronterre&2018} propose a more computationally efficient composite likelihood approach for the case where the positional errors are due to geomasking. 

Intensity estimates may also be adjusted for errors by including relevant covariate information in the model for the intensity. For example, to account for study site selection bias, \cite{Renner&2015} include distance from main roads and distance from urban areas as predictors for modeling the density of {\em Eucalyptus sparsifolia} trees based on presence records from incidental sightings of the species.

\subsection{Perturbed summary statistics}

If the locations are measured with error, the observed locations can be regarded as random perturbations of the underlying correct locations. The summary statistics, such as Ripley's $K$ function, are not the same for the perturbed and correct processes since typically, perturbation makes the process more Poisson like. \cite{Diggle1993} derives the relationship between the perturbed $K$ and original $K$ functions when the (radially symmetric) disturbance model is known. Let $g$ denote the convolution of two independent perturbations. Then the $K$-function of the perturbed pattern becomes
$$
K (r) = \frac{2 \pi}{\lambda} \int_0^\infty R(r,u) u\lambda_2(u)du,
$$
where 
$$
R(r,x) = \int_{\left\|u\right\| \leq r} g(x-u)du
$$
is the probability that a point at $x$ gets perturbed to some point in the region $\left\|u\right\| \leq r$. We see that the result of the perturbation is to smear out the original $K$-function.

The impact of measurement error on the nearest neighbor distribution function $G$ and the empty space statistic $F$ 
is studied and applied to tropical forest data in \cite{BarHenEtal2013}. The unobserved true point process is assumed to be stationary and the perturbation model as in \cite{ChakratobortyGelfand2010}. The summary statistics for $X$ and the distribution of the noise $\epsilon$ are estimated from the observed $Y$ by using the influence of simulated perturbations on the observed point process. 

Using a Bayesian framework, \cite{LundEtal1999} suggest posterior estimates for the summary functions $K$, $G$, and $F$ of a true, unknown point pattern when the observed pattern is exposed to thinning, displacement, or a ghost
pattern. They further provide sampling variation in each case. \citet{KuronenEtal2021} estimate parameters of an actual process based on summary statistics of functions robust with respect to errors.

\subsection{Marking the deviance}

\cite{Santi&2021} consider the problem of estimating spatial autoregressive models when a subset of the data is affected by coarsening, a common result of failures in automatic geocoding. Specifically, a marked point process is employed to simultaneously model the stochastic process and the coarsening mechanism. A maximization of a doubly-marginalized likelihood function of the marked point process then marginalizes out the effects of coarsening. A numerical optimization method is used to ensure computational feasibility. 

\section{Recording and measurement errors}

We now move from a general discussion of modeling frameworks to specific applications, starting with a discussion of various causes of recording and measurement errors. 

\subsection{Microscopy data}
\label{subsection:microscopy}
In our context, microscopy data refers to point patterns that are derived from analyzing objects in microscopy images. Here, we give three examples of microscopy data that can be analyzed by using point processes: confocal microscopy images of epidermal nerve fibers, scanning transmission electron micrograph (STEM) images of silica gel, and photoactivated localization microscopy (PALM) images of proteins in biological cells. Depending on the microscopy technique used, different types of complications may occur. 

In the nerve fiber example, the points where the nerve trees enter the epidermis (outermost layer of the skin), and the points where the nerve fibers terminate, have been analyzed as point patterns in several studies in 2D and 3D, see, e.g., \cite{OlsboEtal2013}, \cite{Andersson2016} and \cite{Konstantinou2021}. The nerve fibers can be followed from one $xy$-image to another but the accuracy of the point locations depends on the chosen resolution, which is typically lower in the $z$-direction than in the $xy$-plane. In the data mentioned here, the resolution in $xy$ is $0.83\times 0.83$ $\mu$m$^2$ compared to 2 $\mu$m in $z$. So far, this difference has not been taken into account in the analysis of such nerve fiber images.

In 2D STEM images of silica, the intensity response is increasing with increasing mass thickness of silica. Therefore, 3D silica nanoparticle structures can be reconstructed from the 2D images if the mass thickness-intensity function is known or can be estimated. Assuming that the intensity response in monotonically increasing with increasing mass thickness, \cite{Nordin2014} estimate the functional form by using a maximum likelihood approach. Which functional form, e.g., linear or power law, is chosen and how the estimated number of silica nanoparticles are located in the $z$-direction will affect the resulting 3D point pattern. 

Photoactivated localization microscopy (PALM) is a powerful imaging technique for characterization of protein organization in biological cells. Due to the stochastic blinking of the fluorescent probes and some camera discretization effects, a cluster of observations is detected instead of a single protein. \cite{Jensen2022} suggest a model, the independent blinking cluster point process, for such data and a method to recover the ground truth based on images observed with the blinking artifact.

\subsection{Particle counters}
Particle contamination is important to identify in machine oil, in clean rooms such as operating theaters in hospitals, and in air quality, among many such applications. The most common type of errors are counting and identification errors.

\cite{rutherford1910} had developed a method for counting alpha particles in radioactive emissions using an electrometer and photographic film. However, it was tedious to use, and instead they use a scintillation method invented by \cite{regener1909}, where a zinc oxide surface gives off a flash when hit by a particle. The scintillations on a small area of the zinc oxide surface were viewed in a microscope, with time reserved for resting the eyes, and recorded on a paper string by hitting an electric switch. Clearly, there is a possibility of missing a scintillation, and of observing a spurious one. In addition, the time of the record is likely offset by a random period of time between seeing the scintillation and hitting the switch. For a homogeneous Poisson process this is of no consequence, but when the process is non-homogeneous or non-Poisson, there are different consequences. Modern radiation counters are based on the Geiger-M\"{u}ller tube, which has a brief dead time after registering a particle.

\cite{martens1968} discusses measurement errors in devices that count (and size) air pollution particles using light scattering. He lists, in addition to sizing errors, three kinds of counting errors, namely instrumental , statistical, and imaginary errors. The term imaginary errors refers to non-repeatability of measurements. The statistical errors are largely due to incorrect particle sizing, while the instrumental errors also can occur, e.g., when two particles are interpreted as one, or there is inhomogeneity of the viewing volume illumination.

Modern tools for particle counting use laser beams, either measuring particle shadows or laser light scattering. The errors are similar: coincidences where two particles are lined up with the laser beam which only sees one shadow, particles that are too small to cast an interpretable shadow, and the same types of light scattering problems as when using incoherent light. There are a variety of other methods, many mentioned in \cite{zhang2009}.

\subsection{Earthquake locations}
The location of the epicenter of an earthquake is determined from translating the temporal distance between P- ans S-waves into a distance, and then triangulating the distance from all the seismometers recording the earthquake. Errors in timing, which can be due to lack of or imprecise coordination with a timing source \citep{hable2018}, then translate to location errors through the triangulation process. More generally, the arrival times can be written as path integrals through the unknown (or at best partially known) structure of earth. The most common approach is iterated linearization \citep[e.g.,][]{menke1989}. It can, however, be unreliable or unstable in the presence of outliers \citep{husen2003}.

\begin{figure*}
    \centering
    \includegraphics[width=\textwidth]{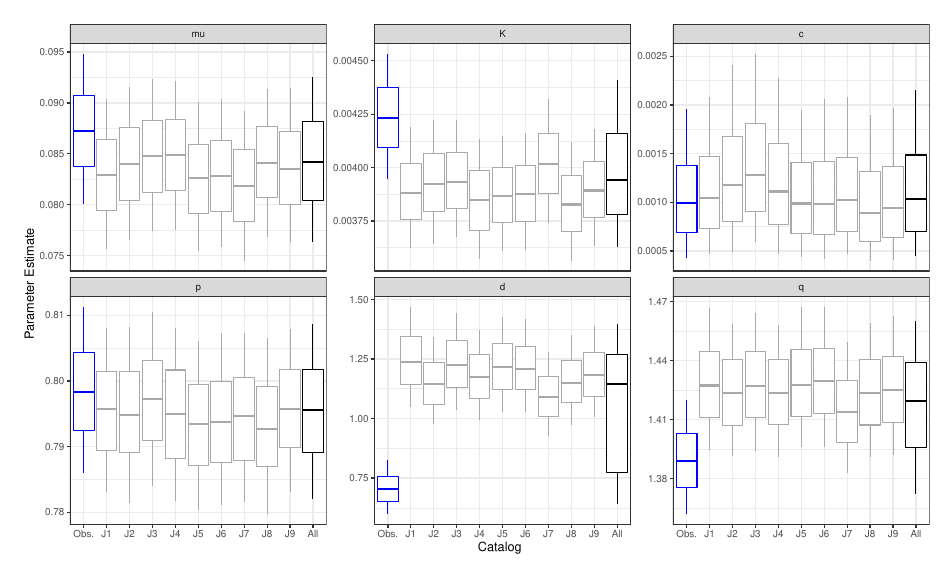}
    \caption{Box plots summarizing posterior densities of ETAS parameters for the reported baseline
catalog (blue), as well as nine jittered versions of this catalog (grey), and for the grand posterior encompassing the reported and 99 jittered catalogs (black). From \cite{Schneider2021}.}
    \label{fig:ETAS}
\end{figure*}

The ETAS (Epidemic Type Aftershock Sequence) model, proposed by \cite{Ogata1998} and extended by \cite{OgataZhuang2006}, is a spatio-temporal point process model for earthquakes. To investigate the effect of measurement error, \cite{Schneider2021} estimates the parameters of an ETAS model to a catalog in the Pacific Northwest, for both the catalog locations as well as jittered locations of the seismic events according to the uncertainties given in the catalog. Figure \ref{fig:ETAS} shows boxplots of original estimates (obtained by MCMC) and nine jittered estimates. We see that the parameters $\mu$ (background rate), $c$ (temporal offset of aftershocks) and $p$ (temporal decay rate of aftershocks) are similar to the estimates from unjittered data, while the estimates of $K$ (aftershock productivity) are biased downwards for the jittered data, while estimates for $d$ (spatial offset of aftershocks) and $q$ (spatial decay rate) are biased upwards, while those for $q$ also are substantially more uncertain.

\subsection{Geocoded data}

Spatial epidemiology is a subfield of epidemiology focused on the study of the spatial distribution of health outcomes. The data analysis is commonly performed by using a spatial point process model to analyze geocoded location data, that is, text-based description of a location---such as an address or the name of a place---transformed to geographic coordinates. Data of this type similarly arises in, e.g., microeconomics and demographic analyses. 

Geocoded data have two different kinds of potential error mechanisms \citep{Fronterre&2018}. Imprecision may be introduced in the geocoding due to incorrect placement of an address along a street segment or the same geocoded location may be assigned to multiple addresses based on a post-code system. Automatic geocoding may also fail for a substantial proportion of the available data and these failures can be spatially clustered \citep{Zimmerman&2010}.  Alternatively, it may not be possible to report precise measurement locations due to issues of confidentiality. Random or deterministic perturbation of the locations is then applied in a process known as geomasking \citep{Armstrong&1999}. See \cite{Jacquez2012} for an overview of the impacts of geocoding positional error in health analysis.

\subsection{Target tracking/Telemetry data}

An alternative to distance sampling in animal ecology is to place a global positioning system (GPS) tracker on the moving object(s) of interest. GPS locations, for example in wildlife tracking \citep{Johnson2013}, have measurement error, and often these errors are ignored. However, there has been some work on looking at the error structure of GPS observations, for example by placing location trackers in different ecological areas. \cite{Williams2012} find that measurement error is higher, by about a factor of two, in areas of dense canopies. Among the error sources are the number of satellites finding the GPS, the temporal resolution of GPS (particularly when following a moving particle, such as an animal), and errors in the positioning or clock of the receiving satellites \citep{Langley1997}. \cite{Ranacher2016} shows that the autocorrelation in GPS locations lead to systematic overestimation of the distance traveled by the object. While many applications to telemetry data don't apply point process approaches, \citet{Arce2023} use a log-Gaussian Cox process model to account for the autocorrelation of subsequent observations in animal tracking data.

\cite{Glennie&2021} combine animal movement data with distance sampling data in order 
to account for animal movement independent of the observer, using a spatial hidden Markov model to approximate the underlying spatial process.

\section{Ghost patterns}

In the applications discussed below, errors in the observation mechanism cause the introduction of an additional ghost pattern in the data, rather than errors in the actual pattern of interest. 

\subsection{Mine fields}
Land mines kill  some  20,000 people worldwide each year. The problem of identifying the location of land mines from satellite, airplane or drone pictures has been considered in the point process literature at least since \cite{rafterydasgupta} and \cite{rafterybyers}. Typically, land mines are located among a clutter of rocks, metal junk, and other features that can be misidentified as mines. \cite{holmes1995} describes a multi-spectral satellite imaging approach to finding land mines, while other methods include infrared thermography (\cite{thanh2009}), 3D voxel radar (\cite{brockner2018}), and polarimetric camera (\cite{connor2009}). 

A point process approach to identification of land mines is based on viewing the observations as a superposition of two point processes: mines and mine-like non-mines. \cite{cressielawson} assume that the two processes are cluster processes with the cluster centers being independent Strauss processes with different parameters, and the cluster points distributed around the cluster centers using a Mat\'{e}rn radial distance function. A hierarchical approach using Markov chain Monte Carlo (MCMC) is taken to estimate the model parameters.

\subsection{Spatial pattern of air bubbles in polar ice}

Polar ice has information on the climate of the past. However, to be able to interpret the ice core records, one has to know how old the ice is. One way to determine the age of ice at different depths is by estimating the deformation of ice based on the air bubbles 
in the ice samples. 

\cite{RedenbachEtal2015} analyze ice samples drilled in Antarctica. The data consist of locations of air bubbles extracted from computer tomographic (CT) images of ice samples imaged inside a cold room at $-15^{\circ}$C .
In addition to the ``real" air bubbles, the ice samples contain some relaxation (extra) bubbles that appear when the ice core is pulled out from the drilling hole. Such bubbles do not give any information on the motion (deformation) of the ice but disturb the directional analysis necessary to gain some information on the deformation. Therefore, each bubble is classified either as real bubble or noise so that the noise bubbles can be removed prior the deformation analysis.

The set-up is very similar to the mine detection problem. However, in this case, it is not necessary that every bubble is classified correctly. It is more important that the spatial pattern of the true bubbles can be reconstructed. For example, in a close pair of bubbles most likely one of them is real and the other one noise and it does not matter which one is classified as real and which one as noise.

\cite{RedenbachEtal2015} suggest the following model for the bubble configuration: The real air bubble pattern is regular and modeled by a Strauss process $Y_1$
with the intensity related parameter $\beta>0$, interaction range $R$, and interaction strength $\gamma$. The noise bubbles are modeled as a Poisson process $Y_0$ with intensity $\lambda_0$. Then, the complete observed point process $Y$ is a superposition of $Y_1$ and  $Y_0$. The parameters $(\lambda_0,\beta,\gamma,R)$ as well as the classification of each point (real or noise) were estimated in a Bayesian framework. In \cite{RedenbachEtal2015}, an MCMC approach is constructed, and a computationally less demanding method based on variational Bayes approximation is suggested in \cite{Rajala2016}. 

Ideally, the points would be divided into two groups, points with high posterior probability (real bubbles) and points with low posterior probability (noise). However, in addition to these two groups, a group of points having intermediate posterior probabilities is found. Typically, the points in this group are $R$-close pairs of points. Some suggestion how to classify the points in this group are suggested in \cite{RedenbachEtal2015} and \cite{Rajala2016}. 
When the model parameters are known the spatial pattern of the true bubbles is estimated quite well. If the parameters need to be estimated simultaneously with the classification, isolated noise bubbles are typically classified as real bubbles, and
as a result, the intensity $\lambda_0$ of the noise bubbles is underestimated and the intensity related parameter $\beta$ of the true bubble process  overestimated (too many real bubbles and too few noise bubbles).

\section{Thinning processes}

In many applications locations of individuals in space and time are recorded, but for practical reasons a specific  observation process has been chosen, which can be seen as a thinning operation on the point pattern. This implies that not all of the points that are truly there have actually been detected.

In some cases, the point pattern has only been partially observed as there are some areas where sampling cannot take place, e.g., for accessibility issues. In other cases the area of interest especially in animal ecology and conservation is simply too big---e.g., an entire ocean---to observe the pattern everywhere. As a result, the
point pattern has only been observed in one or several sub-plots or transects, which have been extensively sampled. This may have been done through what is referred to as plot sampling or transect sampling using areal-video-surveys or drones. It is however of interest to predict into the entire area surrounding the sampled subareas by assuming that some covariates of interest are known within the whole area, with information on point pattern taken from smaller subareas. In these scenarios, the detection probabilities are  1 in the surveyed areas and 0 in the areas that have not been surveyed. In the modeling approach, areas with no points in the surveyed areas and areas with points in the unsurveyed areas have to be distinguished \citep{Williamson2022}.

\subsection{Distance sampling}\label{sec:dist}

Distance sampling is a common method for gathering data on 
population sizes, for example of wildlife populations \cite[e.g.,][]{Buckland&2001, Buckland&2015}. The most common examples for this are line transect sampling, where an observer moves along a linear transect observing the locations of visible objects along the way, as well as their distance from the observer, and point transect sampling, where the observer stands in a location and observes animals within a certain radius, e.g., by listening for birds \citep{Buckland2006}.

In one of the earlier works based on marked point processes, each object (animal) is described by a disc, where the center of the disc is the location of the object and the radius of the disc is the associated sighting distance \citep{Hogmander1991}. An animal is observed if and only if the transect hits the disc connected to it.

On the topic of estimating population sizes based on imperfect detection, \cite{Schmidt&2022} provide a recent overview and a decision support for practitioners to select an appropriate inference approach based on the type of thinning probabilities present in the data collection design. For a probability-based design it is possible to calculate the inclusion probability $p_s$ that a given individual is included in the sample. The detection process of individuals within the sample unit can be divided into $p_p$, the probability that the individual is present within the sample unit, $p_a$, the probability that the individual is available (above ground, above water, vocalizing, etc.) for observation given that it is present, and $p_d$, the probability that an individual is detected given that it is present and available. Commonly, objects close to the observer are more likely to be observed than those further away. 

In some terrains, a subset of the objects might not be visible, e.g., when observing ocean mammals from a water vessel, or due to temporary emigration. This effect is called availability bias and its estimation commonly requires additional information from independent data \citep{LaakeBorchers2004}. Animal movement can cause substantial bias in density estimation from line transect or point surveys. Movement in response to the observer's presence is usually accounted for in the data collection procedure rather than the data modeling procedure \citep{Borchers&1998, Palka&2001}. 

In some cases, detection probabilities vary due to the individuals' behavior or their properties. 
For example, smaller individuals (or groups of  individuals) are more easily missed than larger individuals. In the context of the remote sensing of forests (Section \ref{sec:multitype}), smaller trees are more likely to be missed than larger trees. 

Distance sampling techniques have also been used to estimate the number of trees per hectare and basal area from a terrestrial laser scan \citep{DuceyAstrup2013, astrup_approaches_2014}. In order to correct for the non-detection of trees, which also here is a serious issue, \citet{astrup_approaches_2014} assume that the detectability decreases with increasing distance from the location of the scanner.

Commonly, detection probabilities depend on the distance from the observer and are not known.
\citep{Yuan&2017} consider an approach where the unknown detection function is modeled and its parameters estimated jointly with the underlying point process, i.e.\ as a thinning of a point process accordingly, in particular a log Gaussian Cox process. The intensity is obtained by multiplying the intensity of the log Gaussian Cox process by the detection probability as discussed in Section 3.3. 

In a more general scenario, detection can also depend on properties of the objects represented by the points, e.g.\ the size of an animal or animal group, in other words on the mark in a marked point process. 
For a mark $m$ observed in location $x$ with distribution $p(m|x)$ and the intensity then becomes
$$
\lambda^{\ast} (x, m) = \mathbbm{P}( \text{object at $x$ detected} | x \in X, m ) p(m|x) \lambda (x).
$$

\begin{figure}[!hbpt]
    \centering
    \includegraphics[width=0.6\textwidth]{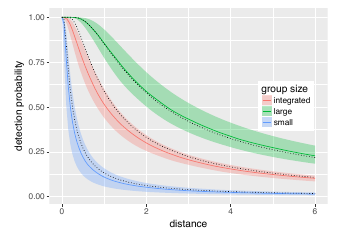}
    \caption{Estimated detection functions for different group sizes.}
    \label{fig:groupsize}
\end{figure}

We consider an example of the pattern formed by the locations of groups of pantropical dolphins (\textit{Stenella attenuata}), with varying sizes of groups to which a model was fitted. Here detection depends on the mark \textit{groupsize}.
Figure \ref{fig:groupsize} shows the estimated half-normal detection functions. Larger groups of animals are easier to detect at larger distances than those at close distances. The model was fitted using the software package \textit{inlabru} which allows the fitting of models using integrated nested Laplace approximations (INLA) without the requirement of a linear relationship between the parameters and the response \citep{bachl2019}.

\subsection{Citizen science}
For presence-only and citizen science data, spatial patterning in presence locations may result from the behavior of observers rather than the behavior of the study species, e.g., due to selection bias in study sites based on accessibility, false absences and false detections \citep{Warton&2013, AltweggNichols2019, SicachaParada&2021}. More generally, biases related to specific data collection mechanisms are mitigated by merging data of different types \citep{Dorazio2014, AltweggNichols2019, Bowler&2019, Renner&2019, Miller&2021, Watson&2021,  Martino2021}.

\section{Multitype deviance}\label{sec:multitype}

Several of the types of discrepancies mentioned above can occur when point patterns of trees are detected using remote sensing such as airborne laser scanning (ALS), aerial photographs (see Section \ref{subsec:deviant}) or hyperspectral data. Depending on how the data are collected and processed, some portion of trees may be missed, e.g., due to large trees blocking the visibility of smaller trees or due to a cluster of trees being erroneously observed as a single large tree. In addition, there may be displacement error related to measurement error or as a consequence of image geometry and lighting conditions. Furthermore, some number of ghost trees can be generated, e.g., by double counting or mistaking branches for tree tops.

\subsection{Airborne laser scanning}

In ALS, laser pulses are emitted from an instrument installed in an airplane and directed to a field plot \cite[]{Vosselman2010}. As a result, a 3D point cloud representing the return locations of the laser pulses is obtained. Individual tree detection (ITD) methods, often based on finding local maxima, are then used to locate tree tops \cite[e.g.,][]{Roussel2020,Roussel2022}. Bayesian methods utilizing certain tree canopy shapes have also been developed \cite[e.g.,][]{LahivaaraEtal2014}. However, no matter how sophisticated the ITD method, errors do occur. Therefore, available ITD methods are not able to reconstruct the spatial structure of the forest \cite[]{PackalenEtal2013}. A few approaches have been developed to correct for undetected trees in ITD methods \citep{KansanenEtal2016, MehtataloEtal2022, KostensaloEtal2023}. Alternatively, it has been proposed to predict the spatial structure of the forest at stand level \cite[]{PippuriEtal2012, HabelEtal2021}.

\citet{Mehtatalo2006} proposes a Horvitz-Thompson-like estimator to correct for trees missed in airborne laser scanning analyses.
The methodology was further developed in \cite{KansanenEtal2016, KansanenEtal2020, MehtataloEtal2022}. In this approach, detected trees are weighted reflecting their detectability utilizing stochastic geometry and canopy size. The core idea is that large trees are always detected and thus are given weight 1. Smaller trees on the other hand
are less likely to be detected and thus they have a weight larger than one, since the detected small trees are only some fraction of the actual number of small trees in the forest.
Then one can estimate, e.g., the stand density with the equation 
\begin{equation*}
\widehat{n}=\sum_{i\in {detected}}\frac{1}{p_i},
\end{equation*}
where $p_i$ is the detection probability for each detected tree.

\citet{KostensaloEtal2023} reconstruct tree patterns from ALS data. Optimized ITD approaches detect only the largest, dominant trees (and trees in sparse plots) and miss approximately 40\% of the trees. In addition, there are about 10\% false discoveries. Clustering is also problematic for ITD, as clustered trees are often detected as a single tree. 

Once the optimized ITD is applied for plots with no ground measurements, a model is used to predict the number of missing trees as well as the number of false discoveries. The placement of the missing trees is based on a resampling-type approach, where a detected tree is selected at random, and a simulated compensating tree is based at a random angle from the detected tree at a distance sampled from the nearest neighbor distribution of the ground measurements. Finally, a proportion of the smallest detected trees are removed as false discoveries. The models for the number of missing trees and false discoveries is based on the plots with ground measurements available.

\begin{figure*}
    \centering
    \includegraphics[width=\textwidth]{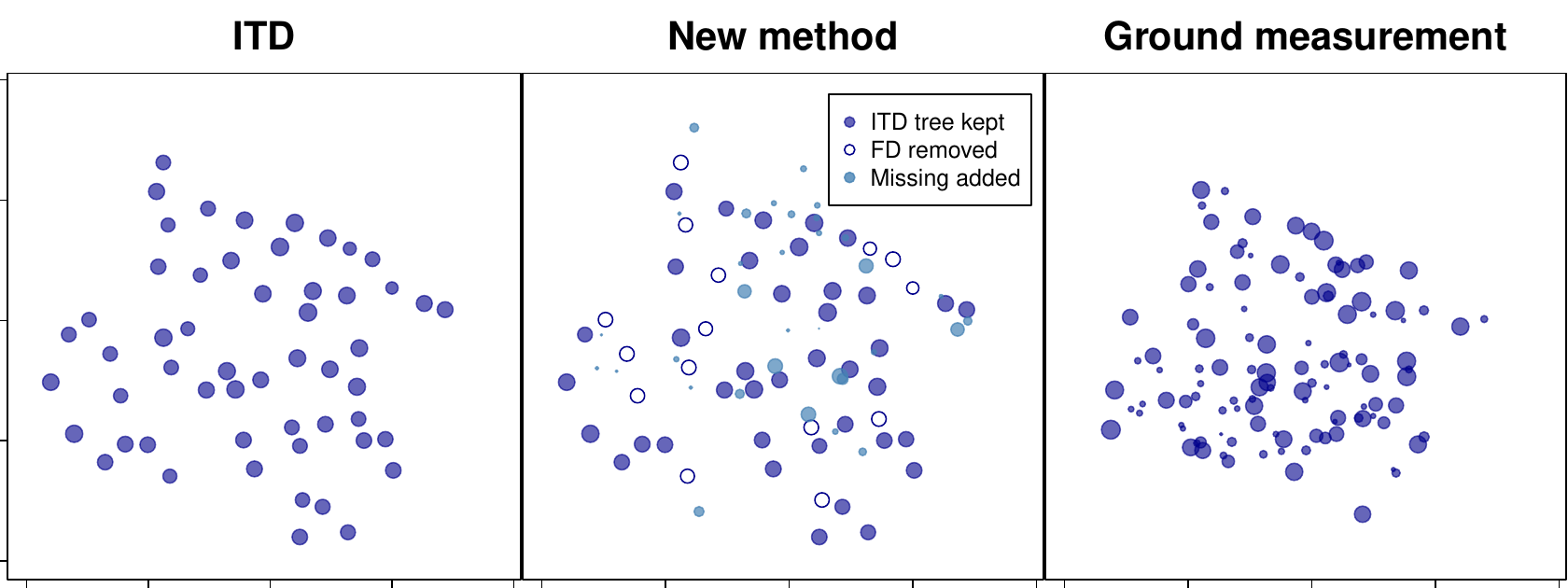}
    \caption{ Left: The pattern of trees detected from ALS data by individual tree detection (ITD). Middle: Tree pattern created using the correction method suggested in \cite{KostensaloEtal2023}. Right: A pattern of trees measured in the field on an area of ca. 400\,m$^2$ in Mikkeli, Finland.
    \label{fig:RS_field_and_ITD_trees}}
\end{figure*}

\begin{figure}
    \centering
    \includegraphics[width=0.45\textwidth]{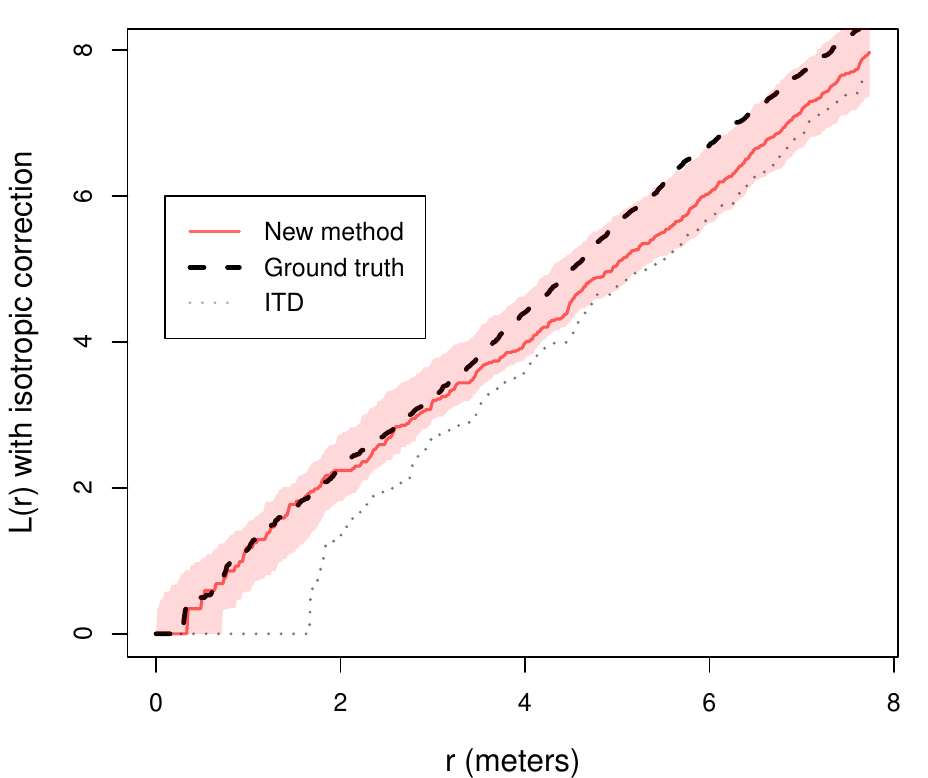}
    \caption{L-function of the three point patterns displayed in Fig.~\ref{fig:RS_field_and_ITD_trees}. The solid red line of the new method corresponds to the particular simulation round shown in the central panel of Fig.~\ref{fig:RS_field_and_ITD_trees} and the shaded area is the 95\% global envelope  estimated from 1000 repeated simulations.
    \label{fig:ITD_res_L_function}}
\end{figure}

Application to one plot in the Mikkeli area in Southern Finland is presented in Fig.~\ref{fig:RS_field_and_ITD_trees}.
The model has been trained with 163 forest plots and the plot is one of the cross-validation test plots.
The $L$-functions with isotropic edge correction of the three point patterns of Fig.~\ref{fig:RS_field_and_ITD_trees} are presented in Fig.~\ref{fig:ITD_res_L_function}, for the proposed (new) method with a 95\% global envelope \citep{MyllymakiEtal2017, MyllymakiMrkvicka2023} constructed from 1000 repeated simulations of the undetected trees. While the $L$-function of the ITD pattern is systematically too regular, the $L$-function of the proposed method is in excellent agreement with the $L$-function based on ground measurements. Thus, while the simulated trees might not be placed exactly where the missing trees are located, the created point pattern captures well the clustering and regularity of the forest plot in addition to the intensity. 

\subsection{Single terrestrial laser scanning}

Terrestrial laser scanning (TLS) is based on a similar principle as ALS, but the laser pulses are emitted from a device that has a fixed location in the field \cite[]{Vosselman2010}. If the forest is scanned from a single location, some of the trees are covered by the ones closer to the scanner and are not detected or are only partially detected. Multiple scans can be used to remove or almost remove the problem of occlusion of trees in a fixed area, but still the extraction of the locations of trees and tree characteristics remains challenging \cite[]{PitkanenEtal2019}. Typically, TLS data have been used to estimate only first-order characteristics of tree data such as density and total basal area \cite[]{KuronenEtal2019, KansanenEtal2020, DuceyAstrup2013} as well as to derive characteristics of trees \cite[]{PitkanenEtal2021}. 

\cite{KuronenEtal2019} construct estimators for forest characteristics such as tree density or basal area based on the visible area of the scanner. They assume that the tree trunks and the cross-sections of tree trunks are perfect circles (no understory vegetation or tree branches affecting the visibility).  To estimate the total basal area, the area of each tree is multiplied by a detection function which is positive only for the detected trees. In addition to the location and size of the tree, the detection function depends on the whole marked point pattern of the trees since the other trees affect whether a tree is detected. When the tree pattern is a realization of a Poisson process, the estimator works quite well but the choice of detection function affects the results. The best results were obtained with the detector, where a tree is detected if any part of it is detected. If the tree pattern is regular, the basal area is estimated larger than it is and this positive bias increases with the degree of regularity. If the tree pattern is clustered, the bias tends to be negative and its absolute value increases with the degree of clustering.
\citet{KansanenEtal2020} propose an alternative estimator using sampling theory instead of point processes. Their estimator is slightly more accurate and has an analytic formula for variance.

\section{Outliers}

An outlier in a spatial point pattern is a point (or points) that is different from the other points in some way. For example, it can be an isolated point or a pair of points that are much closer together than any other pair of points.  
Different methods for goodness-of-fit checking of a particular point process model and finding outliers have been presented in the literature. \cite{Stoyan1991} introduce exponential energy marks that can be used to detect outliers. The points $x_i$, $i=1,\ldots,n$ of a point process $X$ are marked by $m_i=1/\lambda(x_i|\mathbf{x}{\setminus{\{x_i\}}})$, where $\lambda(\cdot|\cdot)$ is the Papangelou conditional intensity. Extreme values of the estimated mark may indicate outliers (with respect to the underlying model). A deviance residual for heterogeneous Poisson processes is defined by \cite{Lawson1993}. The resulting local intensity estimate can be used to find outlying points. Using the Papangelou conditional intensity, \cite{Baddeley2005} introduce residuals for spatial point processes that correspond to the usual residuals for Poisson log-linear regression. The residuals can be used to find outliers, for example isolated points or points with close neighbors. 
\citet{BaddeleyEtal2013} define diagnostics for detecting points with unusual prediction values or points having extraordinarily large influence on the model parameters.
In addition to these model-based approaches, nonparametric functions such as distance to the nearest neighbor can be used  to find outliers whether they are isolated points or close pairs of points.

\begin{figure}[!t]
    \centering
    \includegraphics[width=0.45\textwidth]{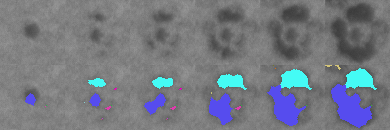}
    \caption{Six frames of the sweat video with an extraordinarily close pair of detected points (top row) and the result of the image segmentation (bottom row), where the different colors indicate different identified sweat glands. 
    \label{fig:spot1}}
\end{figure}

\cite{KuronenEtal2021} regard sweat gland patterns extracted from a video as realizations of spatial point processes. In the patterns observed in the end of the videos, there are some unexpected close pairs of points (gland centers). A closer inspection reveals that some of the sweat spots are detected as two nearby spots due to the several image analysis steps that are needed to extract the point patterns from the videos. By following some of the videos frame by frame, they conclude that some nearby pairs of spots are incorrectly recorded as two glands. An example of such an erroneous pair of sweat glands is given in Figure \ref{fig:spot1}. To avoid the time consuming manual segmentation of the videos, \citep{KuronenEtal2021} regard the falsely detected extra spots as noise 
and introduce two ways to handle the noise. In the first approach, the observed pattern is modeled as a realization of a sequential point process, namely a mixture of a soft-core model for the true sweat gland locations and a uniform distribution for the noise points. 
In the second approach, the underlying unobserved true sweat gland pattern is generated first by using a disturbed simple sequential inhibition process which is then randomly thinned to obtain the observed pattern. The parameters are estimated using an approximate Bayesian computation algorithm \citep{Beumont2010} applied to  summary statistics that are robust to the noise points. 

\begin{figure*}
    \centering
    \includegraphics[width=0.47\textwidth]{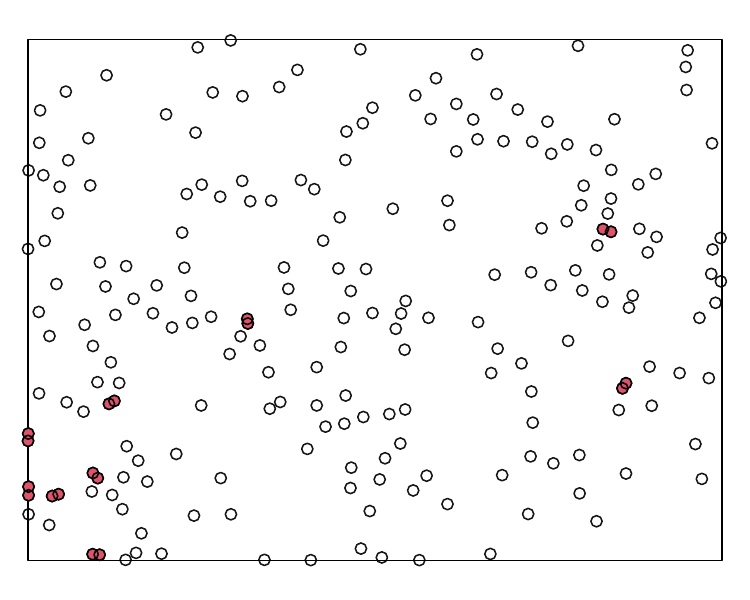}\includegraphics[width=0.47\textwidth]{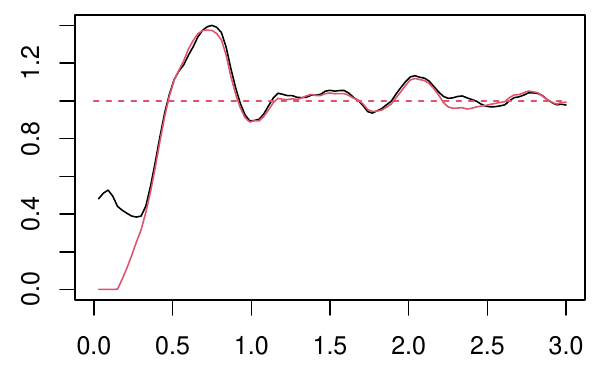}
    \caption{Left: The detected sweat gland locations in a window of size $17.5 \times 13 \text{ mm}^2$. The close pairs of gland locations identified by using the nearest neighbor distances are shown as red discs. Right: The pair-correlation function of the original pattern (black) and the pattern, where one of the points in each outlier pair has been removed (red).
    \label{fig:subject203}}
\end{figure*}

\begin{figure}
    \centering
    \includegraphics[width=0.45\textwidth]{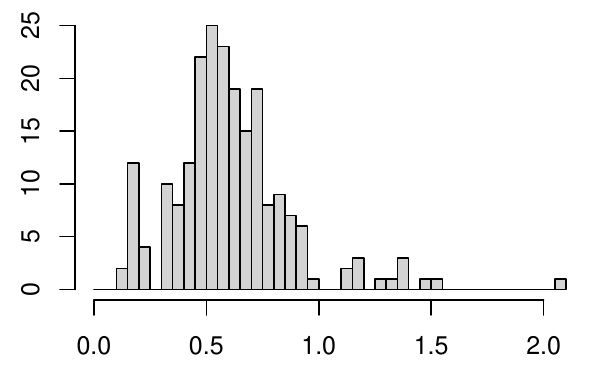}
    \caption{Histogram of the nearest neighbor distances of the original pattern in Figure \ref{fig:subject203}.
    \label{fig:nndist}}
\end{figure}

To complement the analysis in \citep{KuronenEtal2021}, we further consider one of the sweat gland patterns 
(see Figure \ref{fig:subject203}, left) as an example of a pattern with outlying points, or outlying point pairs. 
For each point, we compute the distance to its nearest neighbor and as a result identify eight extraordinarily close pairs of points (the first three bars of the histogram in Figure \ref{fig:nndist}). 
It turns out that these outlier pairs are incorrectly identified sweat gland locations as shown in Figure \ref{fig:spot1}. 
To see how these outlier pairs affect the analysis of the point pattern, we randomly remove one point in each close pair (red discs of Figure \ref{fig:subject203}, left) and estimate the pair correlation function for the original and modified patterns. The behavior at small distances is clearly different in the patterns with and without the outliers (Figure \ref{fig:subject203}, right). 

Then, we investigate how the outlier pairs affect the parameter estimates of a model. Motivated by \cite{KuronenEtal2021}, we use a softcore Gibbs point process model with the density function
\begin{equation*}
f(\mathbf{x}) \propto \beta ^{n(\mathbf{x})}\exp\left(-\sum_{i<j}\left(\frac{\sigma}{\|x_i - x_j\|}\right)^{2/\kappa}\right),
\end{equation*}
where 
$\beta>0$ is an intensity related parameter, and $\sigma>0$ and $\kappa\in (0,1)$ interaction parameters. Large values of $\sigma$ and small values of $\kappa$ indicate strong interaction. 
The model is fitted to the original and modified patterns using spatstat \citep{Baddeley&2015}. 
Especially, the interaction parameter $\sigma$ is affected by the erroneous points (see Table \ref{tab:sweat_est}). 

\begin{table}
\centering
\caption{Pseudo-likelihood estimates of the parameters (95\% confidence intervals) for softcore Gibbs process with all points and with outliers removed. $\kappa=0.2$ was fixed.}
    
    \begin{tabular}{c|c|c|c}
            & $\beta$ & $\sigma$ & $\kappa$ \\ \hline
        All points & 0.95 (0.83, 1.09) & 0.10 (0, 0.13) & 0.2 \\
        Outliers removed & 1.11 (0.95, 1.29) & 0.30 (0.23, 0.32) & 0.2 
    \end{tabular}
    
    \label{tab:sweat_est}
\end{table}

\begin{figure}
   \centering
    \includegraphics[width=0.45\textwidth]{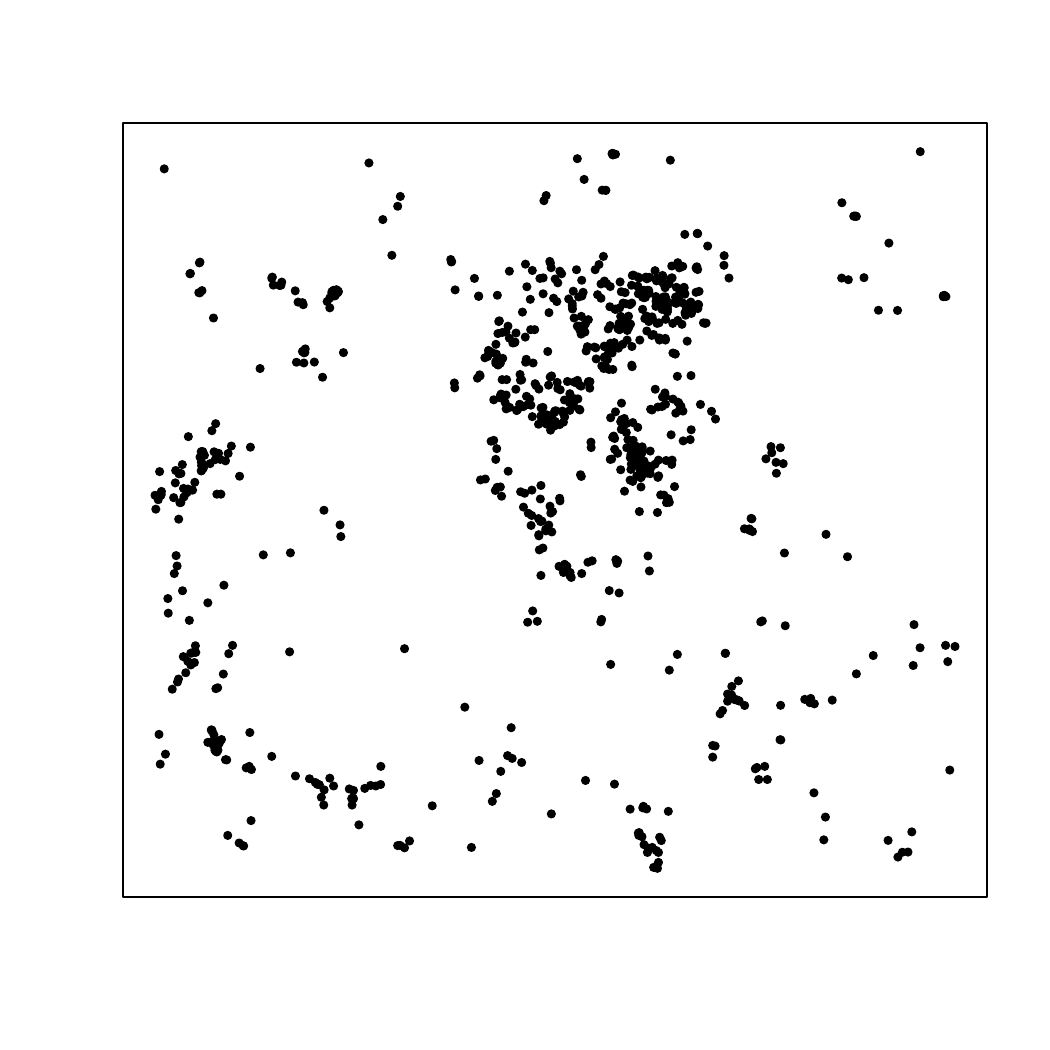}
  \caption{Norway spruce seedlings in a forest stand in Finland.}
\label{outlyingcluster}
\end{figure}

In addition to outlying points or pairs of points, there can be an outlying cluster in a point pattern. 
\cite{KuronenEtal2022} study patterns of Norway spruce seedlings. Figure \ref{outlyingcluster} shows one of the seedling patterns. There is tendency of seedlings to form small clusters. However, there is one extraordinarily large cluster in the middle of the plot. Such clusters are probably not errors (like in PALM data in Section \ref{subsection:microscopy}) but real, although rare, observations. Therefore, the possibility of having such outlying clusters should be included in the model. One possibility would be to model the seedling data by a cluster model where the expected cluster size varies. In the simplest case, the expected number of points in a cluster can take two values, small and large, the large one having a small probability to occur. In general, some information on the underlying, e.g., soil properties would be essential for determining where such clusters are likely to appear. Construction of such models is left for future work.

Slightly related to outlying clusters, \cite{Coeurjolly2017} introduce a median based estimation approach to estimate the intensity of stationary point processes which is robust to outlying areas in the data. An outlying area can be an area without any or with very few points or an area with many points (an outlying cluster) compared to the remaining area. 

\begin{figure}
\centering
  \includegraphics[scale=1]{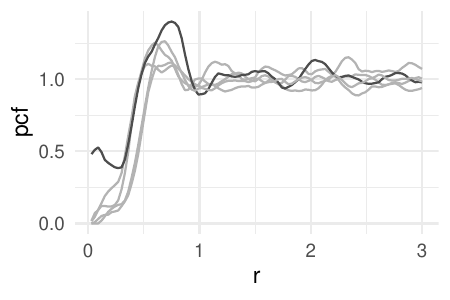}
  \caption{Pair-correlation functions of activated sweat gland patterns from five healthy subjects.}
\label{outlyingpattern}
\end{figure}

Finally, if repetitions are available, one of the patterns can be different from the others and considered an outlying pattern. \cite{KuronenEtal2021} analyze 15 sweat gland videos of which five are from healthy subjects. The estimated pair-correlation functions for the healthy patterns are shown in Figure \ref{outlyingpattern}. One of the patterns (black), also shown in \ref{fig:subject203}, shows a different behavior than the other patterns with more points within a short distance from each other. As pointed out earlier, some close pairs of points in this outlying pattern are incorrectly recorded as two sweat gland locations and by removing such erroneous points, the point pattern is no longer an outlying pattern, see Figure \ref{outlyingpattern} and Figure \ref{fig:subject203} (right). However, if no errors are found, the possibility of outlying patterns should be included in the modeling approach.

\section{Discussion}

Much of the work reviewed in this paper is relatively old, and the computational approaches used may no longer be available. However, it is our experience that unless software is made publicly available, new methodology is very unlikely to be implemented by others. Software for some of the analyses discussed above are available in a GitHub repository\footnote{\url{https://github.com/PointProcess/PointPatternDeviations}}. There are many computer packages dealing with point pattern analysis, but few, if any, explicitly allow errors in the locations of points. We would expect package developers to at least include simple approaches, such as second order estimates for points with i.i.d. location errors.

For a thinned point process, the thinning probability may be changing over space. This may, e.g., be due to varying sampling effort or spatially varying underreporting. A point pattern may have a gap where an unobserved process has thinned the points. A temporal example is given by \cite{LeeBrillinger}, who studied historical Chinese earthquakes, where the probability of an earthquake being recorded depends on which imperial dynasty rules, and the value the dynasty sets on recording historical events. Their approach is parametric, while \cite{GuttorpThompson} estimated the observation probability by smoothing the counting process. Both papers were analyzing the autointensity function, a common second order parameter function for temporal point processes.

Varying thinning probabilities are common in citizen science data, and can sometimes be modeled using covariates such as survey data or data on other species, e.g., \cite{Sicacha-Parada2021}. In the case where a point pattern has gaps due to a not completely known process, one possibility is to assume two overlapping random fields, one with (almost) zero intensity which operates where the empty areas are; this would also work if the deletion process is not deleting all the points, for example using a variant of the approach in \cite{Jones-Todd2018}.

 The availability of image data has increased rapidly in recent years due to advances in satellite and drone technology, to the point where the large data quantities require the use of automated detection algorithms \citep{Eckerstorfer}. In some applications, see e.g. \cite{Eckerstorfer} and \cite{salberg2015detection}, the algorithms have a high detection while they also tend to suffer from high false detection rates, in effect generating ghost points, which must be accounted for in the subsequent data analysis. An appropriate modeling of the detection error can provide valuable input for improving the detection algorithms, calling for a direct synergy between the machine learning detection techniques and the statistical models.

There is currently no systematic way in the literature to deal with point pattern deviations. This review illustrates many of the aspects that would need to be included in such an approach. We are, however, still far from being able to produce an all-encompassing approach.

\section*{Acknowledgements}

JK, MK, and MM were financially supported by the European Union---NextGenerationEU in the Academy of Finland project (Grant number 348154) under Academy of Finland flagship ecosystem for Forest-Human-Machine Interplay---Building Resilience, Redefining Value Networks and Enabling Meaningful Experiences (UNITE) (Grant numbers 337655 and 357909) and AS by Wilhelm and Martina Lundgren's science foundation. We thank Adam Loavenbruck (University of Minnesota) for providing the sweat gland data, Max Schneider (USGS) for allowing us to use a plot out of his dissertation, as well as Sauli Valkonen (Luke) for providing the Norway spruce seedlings data (ERIKA), and Hilkka Ollikainen and Juhani Korhonen for measuring the plots.

\newcommand{\SortNoop}[1]{}

\end{document}